\documentclass[%
 aip,
 amsmath,amssymb,
 reprint,%
]{revtex4-1}

\usepackage{graphicx}
\usepackage{dcolumn}
\usepackage{bm}

\usepackage[utf8]{inputenc}
\usepackage[T1]{fontenc}
\usepackage{mathptmx}
\usepackage{multirow}
\usepackage{mathrsfs}

\usepackage{lineno,hyperref}
\usepackage{graphicx}
\usepackage{subfigure}
\usepackage{booktabs}
\usepackage{epstopdf}
\usepackage{epsfig}
\usepackage{color}
\usepackage{framed}
\usepackage{algorithm}
\usepackage{algorithmic}
\usepackage{array}
\usepackage{makecell}

\begin{document}

\preprint{AIP/123-QED}

\title[Adversarial Attacks to Scale-Free Networks]{Adversarial Attacks to Scale-Free Networks: Testing the Robustness of Physical Criteria}

\author{Qi Xuan}
 \email{xuanqi@zjut.edu.cn}
 
\author{Yalu Shan}%
 \email{1223071127@qq.com}

\author{Jinhuan Wang}
 \email{JinhuanWang@zjut.edu.cn}

\author{Zhongyuan Ruan}
 \email{zyruan@zjut.edu.cn}
 \affiliation{Institute of Cyberspace Security, Zhejiang University of Technology, Hangzhou 310023, China.}
 \affiliation{College of Information Engineering, Zhejiang University of Technology, Hangzhou 310023, China.}
 
\author{Guanrong Chen}
 \email{eegchen@cityu.edu.hk}
 \affiliation{Department of Electrical Engineering, City University of Hong Kong, Hong Kong SAR, China.}

\date{\today}

\begin{abstract}
Adversarial attacks have been alerting the artificial intelligence community recently, since many machine learning algorithms were found vulnerable to malicious attacks. This paper studies adversarial attacks to scale-free networks to test their robustness in terms of statistical measures. In addition to the well-known random link rewiring (RLR) attack, two heuristic attacks are formulated and simulated: degree-addition-based link rewiring (DALR) and degree-interval-based link rewiring (DILR). These three strategies are applied to attack a number of strong scale-free networks of various sizes generated from the Barab\'asi-Albert model. It is found that both DALR and DILR are more effective than RLR, in the sense that rewiring a smaller number of links can succeed in the same attack. However, DILR is as concealed as RLR in the sense that they both are constructed by introducing a relatively small number of changes on several typical structural properties such as average shortest path-length, average clustering coefficient, and average diagonal distance. The results of this paper suggest that to classify a network to be scale-free has to be very careful from the viewpoint of adversarial attack effects. 
\end{abstract}

\pacs{}

\maketitle

%

\section{\label{sec:introduction}Introduction}

Scale-free networks are ubiquitous in nature and society, from email networks~\cite{ebel2002scale} to cell networks~\cite{albert2005scale}, from World-Wide Web~\cite{albert1999internet} to social networks~\cite{xuan2019subgraph}, and beyond. In a scale-free network, few nodes have large numbers of connected links, exhibiting remarkable heterogeneity in node degrees. This special feature makes them be highly robust against random attacks but extremely vulnerable to targeted attacks, which is known as the \emph{Achilles Heel} effect.

Since the influential report on scale-free networks~\cite{barabasi1999emergence}, referred to as the Barab{\'a}si-Albert (BA) network model, there have been a lot of studies \cite{barabasi1999mean,molontay2020}. Generally, a network is considered to be scale free if its degree distribution follows a power-law form, i.e., $p(k)\sim k^{-\alpha}$, where $k$ is the node degree and $\alpha$ is the power-law exponent. In some cases, the definition is stricter. For example, it may require that the power-law exponent satisfies $\alpha\in(2,3)$ or its generation mechanism has a preferential attachment operation~\cite{dorogovtsev2002evolution}. In some other cases, the definition is broader. For example, it may only require the upper tail of the degree distribution curve to satisfy the power-law form~\cite{willinger2009mathematics}, or its log-log curve is nearly straight. The discussions on the definition of a scale-free network has attracted considerable attention since the early 2018, when Broido and Clauset~\cite{broido2019scale} proposed a classification method to estimate the strength of the scale-free attribute of a network. They tested nearly 1,000 real networks and concluded that scale-free networks are rare. Yet, Barab{\'a}si~\cite{barabasi2018love} believes that there are some deficiencies in their preprocessing of the real network data: when transforming a network to several simple degree sequences, the weights for various degree sequences should be different. Voitalov et al.~\cite{voitalov2018scale} pointed out that if the fitted model is a pure power-law model, there is no question that scale-free networks are rare, but in real life, due to the existence of noise, sampling or other processing factors, it is not possible to obtain a pure power-law distribution. Coincidentally, in machine learning, there are many recent studies showing that it is quite possible to output an erroneous result when some slight disturbance such as noise is applied to the input data, which is called adversarial attack~\cite{szegedy2013intriguing}. Consequently, a great deal of interest is aroused to revisit the notion of adversarial attacks.

Noise plays an important role in many aspects of data analysis. Since deep learning is widely used in computer vision~\cite{xuan2017automatic,xuan2018multiview}, there are a number of studies on algorithm robustness~\cite{papernot2016limitations,su2019one,moosavi2016deepfool,xie2017adversarial}. It was found that changing a few pixels in an image could make the classification result totally wrong or different. The reason is that the feature vector of the image will change when a pixel is modified, which can fool many deep learning algorithms. For example, Su et al.~\cite{su2019one} changed only one pixel in an image, they were able to destroy many intrinsic properties of the image and fool a deep neural network. Besides, there are many other ways to add such disturbances to an image, e.g., FGSM~\cite{goodfellow2014explaining}, ILCM~\cite{kurakin2016adversarial}, DEEPFOOL~\cite{moosavi2016deepfool}, and so on. Most of these methods can only generate a specific disturbance for a specific image, instead Moosavi-Dezfooli et al.~\cite{moosavi2017universal} created general perturbations for a bunch of images, making the situations even worse, since such disturbances are not easily identifiable by human.

Besides computer vision, in the area of network science it was also found that simple purposeful modifications to an original network, i.e. by rewiring links, adding/deleting nodes or changing nodes' attributes, can significantly change the network properties and thus greatly disturb the graph algorithms~\cite{dai2018adversarial}. Recently, a number of strategies were proposed to attack link prediction~\cite{fu2018link,chen2019lstm}, node classification~\cite{wang2018attack}, and community detection~\cite{chen2019ga}. For instance, Yu et al.~\cite{yu2019target} proposed heuristic and evolutionary algorithms to protect targeted links from link-prediction-based attacks. Z{\"u}gner et al. ~\cite{zugner2018adversarial} considered attacking node classification, and introduced NETTACK based on a graph convolutional network (GCN) to generate adversarial attack iteratively. Chen et al.~\cite{chen2019ga} used genetic algorithm (GA)-based $Q$-attack to destroy the community structure of a network, where the modularity $Q$ is used to design the fitness function.

On the other hand, $p$-value is dominant in statistical analysis, which however has been questioned before~\cite{evans1988end}. Today, the $p$-value measure has been re-examined and advanced to a new climax~\cite{amrhein2019scientists}. More than 800 reports pointed out that the sample size has a great impact on the $p$-value. By adding or subtracting some data, which can be considered as noise, one will get a different result compared with the original one. This explains why most experimental results obtained by using $p$-values are difficult to reproduce.

The reality is, unfortunately, what you see from the data may not mean what they really are. In other words, the algorithms and models developed based on real data could be vulnerable to tiny noise and purposefully designed noise-like perturbations. Therefore, the robustness of machine learning algorithms attracts more and more attention from the Artificial Intelligence (AI) community. The robustness of many physical criteria, especially those based on data, have to be examined to ensure that they are reliable.

Motivated by the above discussions, in this paper, a new type of \emph{adversarial attack} is introduced to the robustness of physical criteria on \emph{scale-free} networks. Specifically, three attack strategies are used to evaluate the robustness of the classification method proposed for scale-free networks~\cite{broido2019scale}. To be convincing, the Barab\'asi-Albert (BA) model is used to generate 100 \emph{strong} scale-free networks of different sizes for testing the attacks in experiments, by rewiring some links of these networks until the Broido-Clauset (BC)~\cite{broido2019scale} classification goes wrong. The results show that the BC classification can be easily fooled by rewiring only a small fraction of links.

\begin{figure*}[!t]
 \centering
  \includegraphics[width=\linewidth]{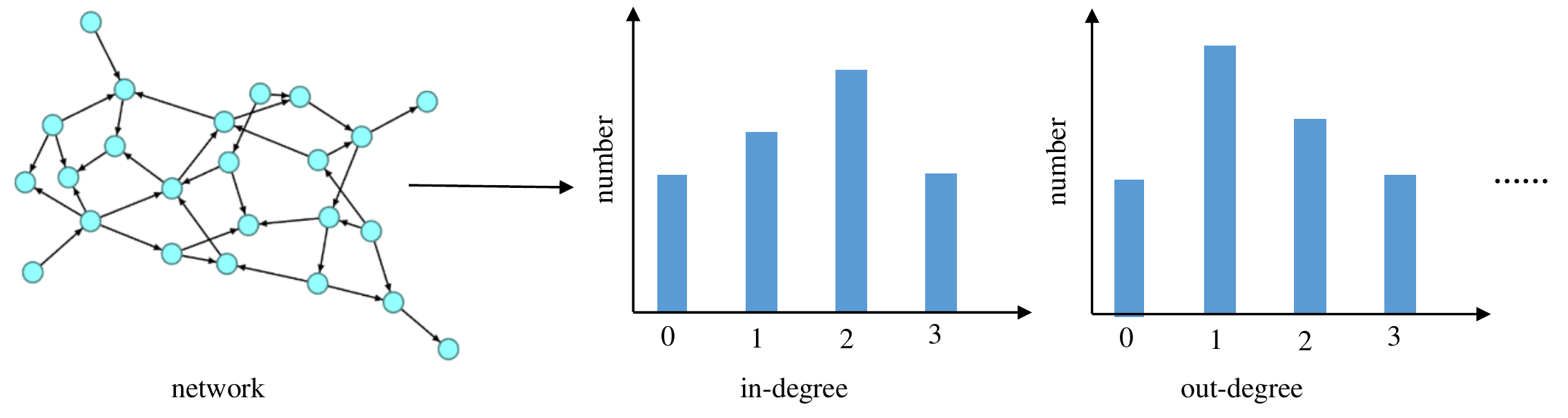}
  \caption{Preprocessing of a real-world network under the BC classification framework. Taking a directed network as an example, the directed attribute is removed and both in-degree and out-degree sequences are obtained.}
  \label{fgpro}
\end{figure*}

The main contributions of this paper are summarized as follows.
\begin{itemize}
\item A new \emph{adversarial attack} is introduced onto the physical criteria of scale-free networks to evaluate the robustness of the scale-free measure.
\item Two heuristic attack strategies, namely \emph{degree-addition-based link rewiring} (DALR) and \emph{degree-interval-based link rewiring} (DILR), are introduced. Several structural metrics are proposed to measure the effectiveness and concealment of the attack strategies.
\item It is found that both DALR and DILR are more effective than random link rewiring (RLR), in terms of rewiring fewer links to successfully attack strong scale-free networks, so that the networks become other types. It is also found that DILR is as concealed as RLR in that they both are constructed by introducing a relatively small number of changes on several typical structural properties such as average shortest path-length, average clustering coefficient, and average diagonal distance. Therefore, the results of this paper suggest that to classify a network to be scale-free has to be very careful from the viewpoint of adversarial attack effects.
\end{itemize}

The rest of the paper is organized as follows. The BC classification of scale-free networks is introduced in Sec.~\ref{sec2}. Three adversarial attack strategies are proposed in Sec.~\ref{method}. Some metrics for attack effectiveness and concealment are discussed with experimental results reported in Sec.~\ref{exp}. Conclusions are drawn in Sec.~\ref{con}.

\section{Classification of scale-free networks}\label{sec2}

Here, the classification of scale-free networks proposed by Broido and Clauset~\cite{broido2019scale} is reviewed, which is referred to as the BC classification below.

\begin{figure}[t]
 \centering
  \includegraphics[scale=0.6]{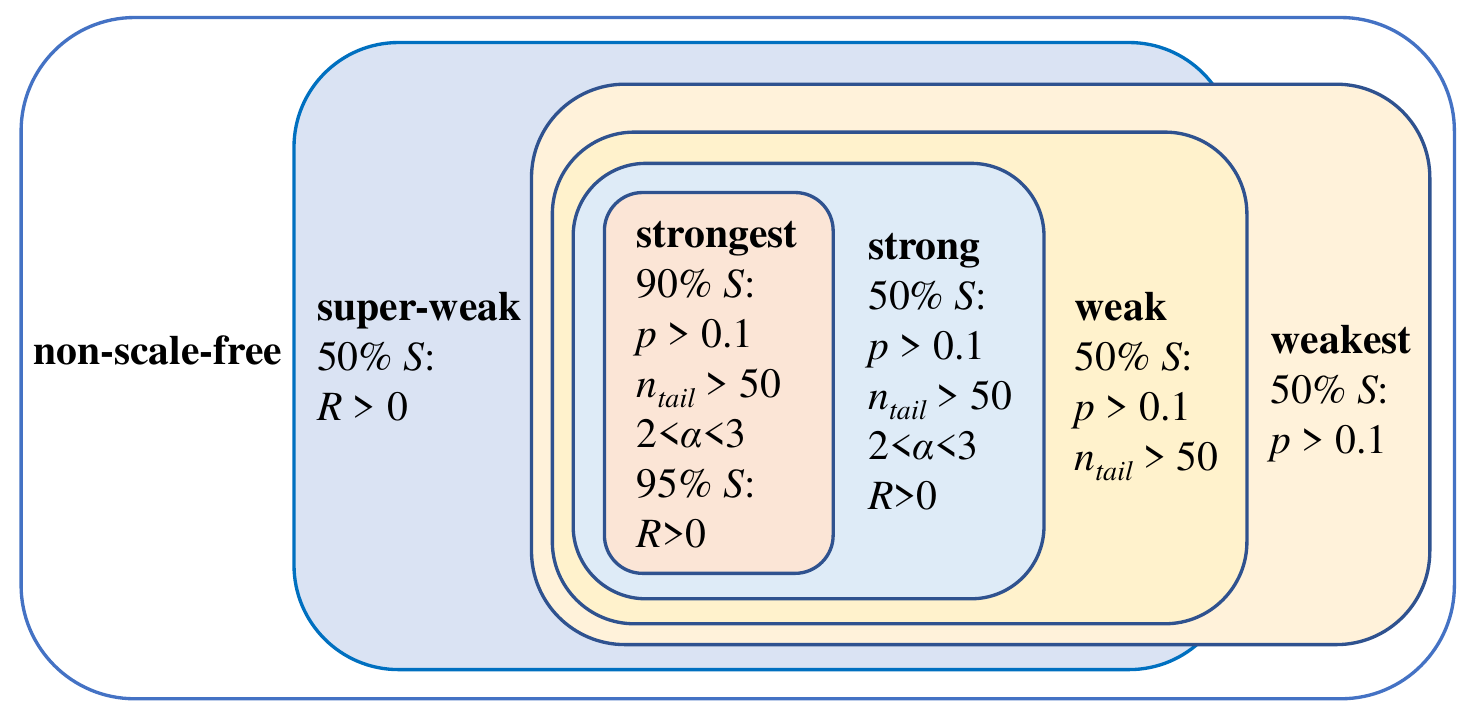}
  \caption{The real-world networks in examination are classified into six categories~\cite{broido2019scale}: strongest, strong, weak, weakest, super-weak, and non-scale-free.}
  \label{fgreal}
\end{figure}

\begin{itemize}
\item \emph{First, preprocessing the real-world network data.} Common analysis typically relies on the \emph{scale-free hypothesis} to determine whether a network is scale-free or not. This hypothesis states that a network is scale-free if its degree distribution follows a power-law form, where only the information of node degree is taken into account. While real-world networks carry many irrelevant attributes such as link weights, connection directions, etc., which make it inconvenient to estimate the strength of the scale-free attribute. Thus, the BC classification suggests to first convert the original network to a set of simple graphs, where one can apply the scale-free hypothesis to each graph. In this way, one can obtain several degree sequences. For example, from a directed network, one can get an in-degree sequence and an out-degree sequence (see Fig.~\ref{fgpro}). To that end, one can put all the degree sequences in a set, $S$.

\item \emph{Second, estimating the strength of scale-free attribute.} Several indicators are used in BC classification to estimate the strength of scale-free attribute for a given network, which are listed in TABLE~\ref{tab:indicators}, where $R$ is one newly proposed here to evaluate the best fit model.

\begin{table}[htbp]
  \centering
  \caption{Indicators for the scale-free attribute in BC classification.}
    \begin{tabular}{cp{23em}}
    \toprule
    \multicolumn{1}{c}{\textbf{Indicators}} & \textbf{Description} \\
    \midrule
    \multicolumn{1}{c}{\multirow{2}[1]{*}{\textit{$\alpha$}}} & Power-law exponent obtained by fitting a power-law degree distribution model. \\
    \midrule
    \multicolumn{1}{c}{\multirow{1}[1]{*}{\textit{$n_{tail}$}}} & Number of tail nodes used for fitting. \\
    \midrule
    \multicolumn{1}{c}{\multirow{2}[1]{*}{\textit{$p$}}} & $p\ge0.1$: Accept the scale-free hypothesis.\\ & $p<0.1$: Reject the scale-free hypothesis. \\
    \midrule
    \multicolumn{1}{c}{\multirow{5}[1]{*}{\textit{$R$}}} & $R>0$: The power-law model is much in favor.\\ & $R=0$: The data do not permit a distinction between (power-law or alternative) models. \\& $R<0$: The alternative model (such as the exponential model) is much in favor. \\
    \bottomrule
    \end{tabular}%
  \label{tab:indicators}%
\end{table}%

\item \emph{Third, classifying the networks.} According to the above two steps, one can determine the strength of the scale-free attribute of a network, based on which, the real networks in examination were divided into six categories: \emph{strongest}, \emph{strong}, \emph{weak}, \emph{weakest}, \emph{super-weak}, and \emph{non-scale-free}, as shown in Fig.~\ref{fgreal}. It can be seen that the strongest, strong, weak, and weakest classes are nested, indicating that the strength of the scale-free attribute becomes weaker gradually. The super-weak class indicates that the networks are extremely weak in scale-free attribute, which only requires that the optimal distribution is in a power-law form. Networks that do not fall into the above five categories are considered as non-scale-free networks.
\end{itemize}

Broido and Clauset~\cite{broido2019scale} analyzed nearly 1,000 networks in different domains, and they found that only a small number of them can be considered as strong scale-free networks. Even the simplex networks generated by the BA model do not entirely belong to the strong category. Thus, they concluded that scale-free networks are rare. This classification method has attracted much attention in the network science community, and triggered wide discussions recently~\cite{barabasi2018love,gerlach2019testing,voitalov2018scale}.

\section{Method}\label{method}

In a computer vision simulation, as an adversarial attack example, tiny perturbations are added into a panda picture. It is very difficult for human to find differences between the attacked picture and the original one. This surprisingly can make the classification algorithm misjudge it to be a gibbon picture with a probability of 99.3\%~\cite{goodfellow2014explaining}.

In this paper, the idea of adversarial attack is applied to fool the BC classification introduced in the previous section. Specifically, by rewiring a few links in the network to alter the values of the indicators, namely making the indicators exceed the limit of the original category, the attack can make the BC classification results go wrong. According to the BC classification algorithm, the difference between the strong category and the weak category is determined by the power-law exponent $\alpha$ and the indicator $R$ in TABLE~\ref{tab:indicators}, and the difference between the weak and weakest categories is determined by $n_{tail}$, the number of fitting nodes. It is emphasized that the power-law exponent $\alpha$ plays an more important role than the indicator $R$ on determining the strong category~\cite{broido2019scale}. Therefore, in this paper, the adversarial attack strategy is designed based on the measures of both power-law exponent $\alpha$ and number of fitting nodes $n_{tail}$.

A network is represented by a graph $G=(V,E)$, where $V$ and $E$ represent the sets of nodes and links in the network, respectively. Let $\Omega$ represent the set of links in the fully-connected network with the same set of nodes $V$. Note that $E$ is a part of set $\Omega$. Define $N=\Omega-E$ as the set of links in the complementary network of $G$, namely those in the fully-connected network but are not in network $G$.

\subsection{Random Link Rewiring}

In the random link rewiring (RLR) scheme, one first randomly selects a link $l_{delete}\in E$ to delete and a link $l_{add}\in N$ to add at each step time, so as to balance the number of links in the network. After an attack, the set $E$ and $N$ become $E'$ and $N'$, respectively:
\begin{equation}
E'=E-l_{delete}+l_{add},
\end{equation}
\begin{equation}
N'=N+l_{delete}-l_{add}.
\end{equation}
Thus, one gets a new \emph{adversarial network} $G'=(V, E')$.

\subsection{Degree Based Link Rewiring}

One of the prominent properties of scale-free networks is that there exist a few nodes with extremely large degrees, called hubs. These hubs are key to the robustness of the networks---they make the scale-free networks be highly tolerant to random attacks while extremely vulnerable to targeted attacks~\cite{albert2000error}, measured by the connectivity or the average shortest path-length of the network. Moreover, they play an important role in some dynamical processes taking place on the networks, like epidemic spreading~\cite{pastor2003epidemics,ruan2012epidemic}, information cascading~\cite{wolfson2009signaling} and synchronization~\cite{wang2002pinning,xuan2017social}.

\textbf{Degree-Addition-based Link Rewiring (DALR).}

Roughly divide the links in a scale-free network into three categories: hub-hub links (links connecting two hubs), hub-normal links (links connecting one hub and one normal node), and normal-normal links (links connecting two normal nodes). Then, DALR is designed based on the following two operators.
\begin{itemize}
    \item \emph{Deleting a hub-hub link.} Since the number of hub-hub links in a scale-free network are generally quite small, this condition can be relaxed as follows: first, design an indicator $d_{(i+j)}$ to measure the degree of node pairs:
    \begin{equation}
    d_{(i+j)}=d_i+d_j,
    \end{equation}
    where $d_i$ and $d_j$ represent the degrees of nodes $v_i$ and $v_j$, respectively. Then, sort the links according to their values of $d_{(i+j)}$, from large to small. The link with the highest ranking will be chosen to remove.
    \item \emph{Adding a normal-normal link.} In order to weaken the scale-free attribute without changing the network density, add a normal-normal link, since this operation can weaken the heterogeneity of the network. Specifically, the nodes are sorted according to their degree from small to large, and the link between the pair of unconnected nodes with the smallest sum of degrees is chosen to connect together.
\end{itemize}

\begin{figure*}[!t]
  \centering
  \includegraphics[width=\linewidth]{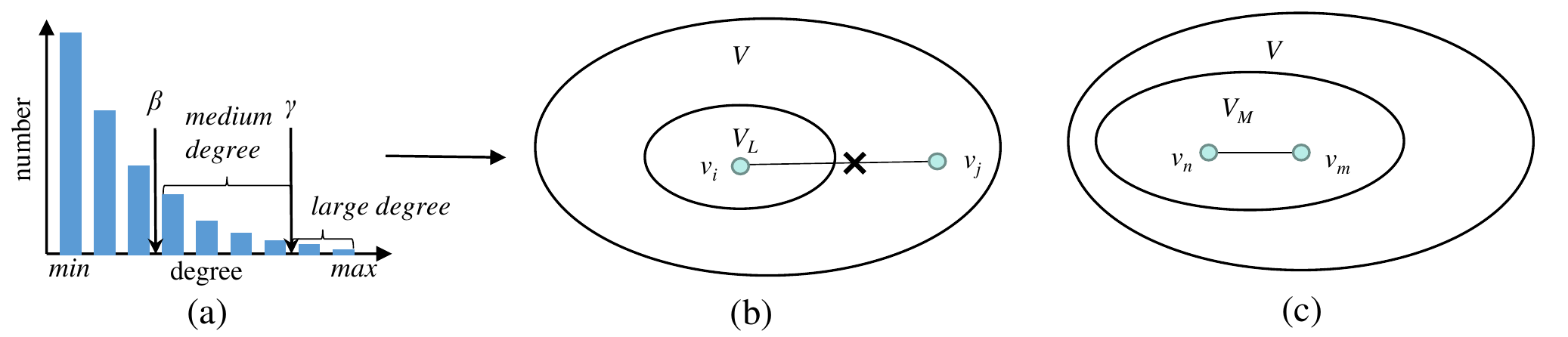}
  \caption{Illustration of DILR attack strategy. (a) Determine the set of large-degree nodes, $V_L$, and the set of medium-degree nodes, $V_M$, according to the degree sequence. (b) Delete the link between two connected hub nodes (note that, here, node $v_j$ has the largest degree among all the neighbors of $v_i$, so it could be either in or not in $V_L$). (c) Add a link between two unconnected nodes of medium degrees.}
  \label{fgHA}
\end{figure*}

\textbf{Degree-Interval-based Link Rewiring (DILR).}

The main indicators of strong and weak categories are the power-law index $\alpha$ and the number of fitting nodes, $n_{tail}$, which are determined by the fitting process. Divide the node distribution into two parts according to the $p$ value: the tail of the distribution used for fitting and the others for another purpose. As is well known, the power-law index $\alpha$ is determined by the (average) slope of the curve; therefore, if the tail is steeper with smaller $n_{tail}$, it will be out of the strong and weak categories much easily.

Based on the above observations, one can roughly divide the nodes into three sets: nodes with large degrees (within $\gamma$=20\%), nodes with medium degrees (between $\gamma$ and $\beta$, with $\beta$ varying from 35\% to 70\%), and nodes with small degrees (the remaining ones), as shown in Fig.~\ref{fgHA}. Denote $V_L$ and $V_M$ the nodes of large degrees and medium degrees, respectively. Then, the following DILR attack strategy is designed.
\begin{itemize}
    \item \emph{Deleting a link between two connected hub nodes.} First, select a node $v_i\in{V_L}$, and then choose a $v_j\in{N_i}$, where $N_i$ is the neighboring set of $v_i$, such that $v_j$ has the largest degree among all neighbors of $v_i$. Then, delete the link between $v_i$ and $v_j$.
    \item \emph{Adding a link between two unconnected nodes of medium degrees.} Randomly select two unconnected nodes from $V_M$, and add a link between them.
\end{itemize}

\section{Experimental Results}\label{exp}

\begin{figure}[!h]
 \centering
  \includegraphics[scale=0.7]{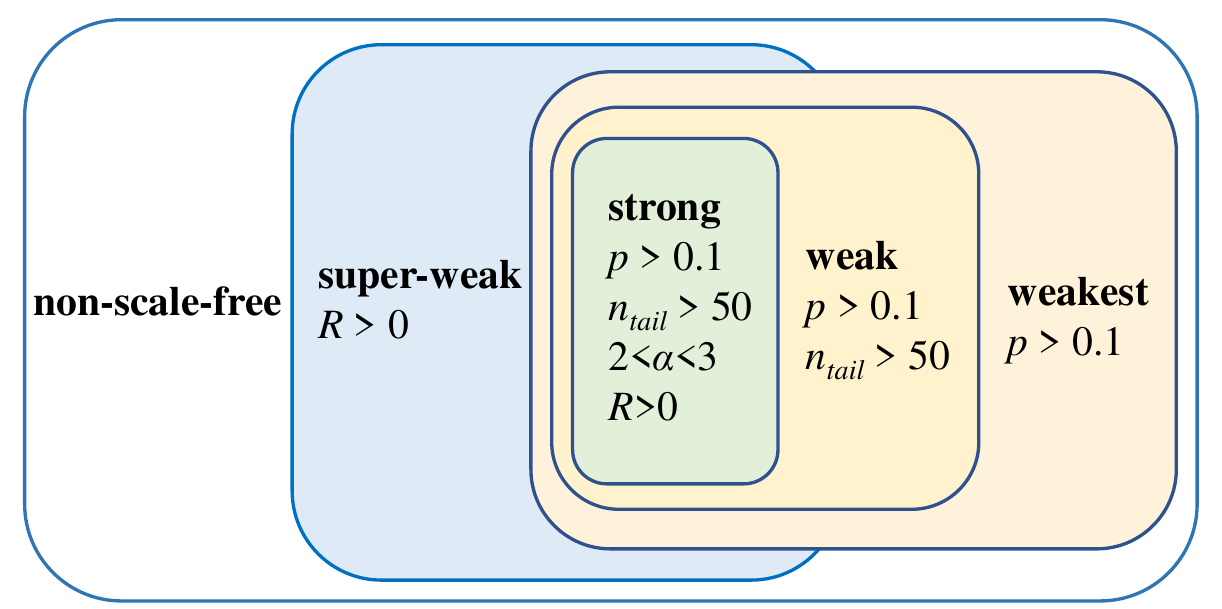}
  \caption{Classification of simplex networks generated by the BA model, in terms of scale-free attribute. Here, strongest and strong categories are merged into one, the strong category, and the other four categories keep the same, as suggested by Broido and Clauset~\cite{broido2019scale}.}
  \label{fgsimple}
\end{figure}

\subsection{Datasets}
There are many models for generating scale-free networks: the BA model~\cite{barabasi1999emergence}, fitness model~\cite{bianconi2001bose}, and local-world evolving network models~\cite{lichen2003,xuan2006growth,xuan2007local,sen2009new}, etc. In this paper, the most well-known BA model is adopted to generate simplex networks. Since all the generated networks are simplex, there will not be the strongest category. In other words, only the following five categories are considered here: \emph{strong}, \emph{weak}, \emph{weakest}, \emph{super-weak}, and \emph{non-scale-free}, where the former three are nested, as shown in Fig.~\ref{fgsimple}.
In order to test the robustness of the BC classification method, the simulation is to attack the BA network of different sizes: $n=500,\, 1000,\, 2000$, respectively, where $n$ is the number of nodes. For simplicity, the number of links, $m$, is set to be $2n$.

In the simulation, 500 networks for each size were generated, and then classified. TABLE~\ref{2} shows the classification results, where it can be seen that most (but not all) of the networks generated by the BA model indeed fall into the strong category regardless of the sizes of the networks. However, there are still some networks that cannot be considered as scale-free networks by the BC classification, specifically more than 10\% of them are considered as super-weak and very few of them are considered as weak. As the network size increases, more networks will be classified into the non-scale-free category.

\subsection{Performance Metrics}

\subsubsection{Metric for effectiveness}

When attacking a model or scheme, one always hopes that the attack could successfully fool the object with a lowest cost, e.g., changing the least number of pixels to fool a computer vision model, or rewiring the least number of links to fool a link prediction or a node classification algorithm. Here, to measure the effectiveness of different adversarial attack strategies, the smallest fraction of rewired links needed to successfully fool the BC classification method~\cite{broido2019scale} is considered, by changing the \emph{strong} category of scale-free networks to another type. The metric is defined by
\begin{equation}
\Delta{M}=\frac{m_R}{m},
\end{equation}
where $m_{R}$ is the smallest number of rewired links to realize a successful attack and $m$ is the total number of links in the whole network. If a (strong) scale-free network is still in the strong category after a large perturbation introduced by the attack, i.e., a large fraction of links are rewired, then it is deemed that the classification algorithm is robust or the attack is less effective.

\begin{table}[!t]
  \centering
  \caption{Classification results of the BA network of different sizes. Most (more than 70\%) but not all of these BA networks fall into the strong category.}
    \begin{tabular}{cccccc}
    \toprule
    \multicolumn{1}{p{1.em}}{\textbf{n}} & \multicolumn{1}{p{2em}}{\textbf{strong}} & \multicolumn{1}{p{2.555em}}{\textbf{weak}} & \multicolumn{1}{p{3.055em}}{\textbf{weakest}} & \multicolumn{1}{p{5.78em}}{\textbf{super-weak}} & \multicolumn{1}{p{7em}}{\textbf{non-scale-free}} \\
    \midrule
    500   & 89.30\% & 0.30\% & 0\%   & 10.30\% & 0\% \\
    1000  & 78\%  & 0\%   & 0\%   & 20.60\% & 1.40\% \\
    2000  & 73.60\% & 0.20\% & 0\%   & 14.60\% & 11.60\% \\
    \bottomrule
    \end{tabular}%
  \label{2}%
\end{table}%

\subsubsection{Metrics for concealment}

Generally, rewiring a network not only may change the scale-free attribute (the objective here), but also may change other structural properties of the network. In this sense, defenders may utilize certain structural properties to discover the adversarial attack.

To quantify the concealment of various adversarial attacks, the following three metrics are proposed for measuring the structural changes introduced by an attack.

\begin{itemize}
    \item \textbf{Relative change of average distance ($\Delta{L}$).} The average distance, or shortest path-length, is defined as
\begin{equation}
L=\frac{2}{n(n-1)}\sum_{i>=j}d_{ij},
\end{equation}
where $n$ is the number of nodes and $d_{ij}$ is the shortest path length between nodes $v_i$ and $v_j$.

The average shortest path length is a most typical global characteristic of a network. Due to the existence of hubs, the average shortest path of a scale-free network is generally shorter than that of a random network~\cite{ugander2011anatomy}. This means that if a network has a strong evidence to be scale-free, its average path-length should always be short.

Here, the focus is on the relative change of $L$ introduced by the attack. Denote by $L_{original}$ and $L_{adversarial}$ the average shortest path-length of the original network and the adversarial network, respectively. Then, the relative change of $L$ is defined as
\begin{equation}
\Delta{L}=\frac{|L_{adversarial}-L_{original}|}{L_{original}}.
\end{equation}

\item \textbf{Relative change of average clustering coefficient ($\Delta{C}$).} Clustering Coefficient~\cite{watts1998collective} is defined as the ratio of the actual number of links among the neighbors of a node to the maximum possible number of links among the neighbors. The average clustering coefficient of a whole network is defined as
\begin{equation}
C=\frac{1}{n}\sum_{i=1}^{n}{\frac{2E_i}{k_i(k_i-1)}},
\end{equation}
where $k_i$ is the number of neighbors of node $v_i$ and $E_i$ is the actual number of links among the neighbors of $v_i$.

The clustering coefficient, as the most typical local property, reflects how densely the neighbors of a node are connected to each other. In scale-free networks, the probability of having a link between two nodes connecting to the same node is relatively low. As a result, scale-free networks generally have small average clustering coefficients.

Similarly, here the focus is on the relative change of $C$ introduced by the attack. Denote by $C_{original}$ and $C_{adversarial}$ the average clustering coefficient of original network and the adversarial network, respectively. Then, the relative change of $C$ is defined as
\begin{equation}
\Delta{C}=\frac{|C_{adversarial}-C_{original}|}{C_{original}}.
\label{DeltaC}
\end{equation}

\item \textbf{Relative change of diagonal distance ($\Delta{D}$).} Diagonal distance~\cite{tsiotas2019detecting}, denoted by $D$, measures the average distance from the main diagonal of nonzero elements in the adjacency matrix of a graph, which is defined as
\[
D=dd(A)=\langle dd \rangle =\frac{1}{n^2} \sum_{(i,j)\in E}{dd_{ij}}
\]
\begin{equation}
= \frac{1}{n^2} \sum_{(i,j)\in E}{\frac{x_{ij} \cdot y_{i,j}}{z_{ij}}}= \frac{1}{\sqrt{2} \cdot n^2} \sum_{(i,j)\in E}{\vert{i-j}\vert},
\end{equation}
where $dd_{ij}$ is the distance of the elements $i$ and $j$ from the main diagonal of the adjacency matrix $A$, $x_{ij}=\vert(i,i)-(i,j)\vert$, $y_{ij}=\vert(j,j)-(i,j)\vert$, $z_{ij}=\sqrt{x_{ij}^2+y_{ij}^2}$, $n$ is the number of network nodes, and $\langle \cdot \rangle$ is the average operation.

It was found~\cite{tsiotas2019detecting} that diagonal distance can be used to detect the dramatic changes in the topology of a network. Here, the relative change of $D$ is used to measure the attack concealment, which is defined as
\begin{equation}
\Delta{D}=\frac{|D_{adversarial}-D_{original}|}{D_{original}},
\end{equation}
where $D_{original}$ and $D_{adversarial}$ are the average diagonal distances of the original network and the adversarial network, respectively.
\end{itemize}

\subsection{Results}

In this study, 100 networks were randomly selected from the strong category in the generated networks of each size $n$ to attack, for $n=500,\, 1000,\, 2000$ respectively, with number of links $m=2n$.

Note that the most important indicators for distinguishing strong and weak categories are the power-law exponent $\alpha$ and the indicator $R$ in TABLE~\ref{tab:indicators}. Therefore, the goal here is to make the power-law exponent $\alpha$ be out of the range of the strong category by using the attack strategies introduced in Sec.~\ref{method}.
An attack is repeated until the network does not meet the strong category requirement. Then, the resultant adversarial network is reclassified.

The results show that such an attack not only changes the power-law exponent $\alpha$, but also changes the indicators $p$ and $n_{tail}$, which makes the adversarial networks be chanced to other categories. However, the probability of falling into each new category is different. To reduce the contingency of the attack results, each network is attacked for 200 times and the mean results are recorded. In the simulation, it was found that 200 times are enough for the probability to converge for the tested networks.

\begin{table*}[htbp]
  \centering
  \caption{The minimum fractions of rewired links ($\Delta{M}$) needed for a successful attack, for different attack strategies on networks of different sizes. Here, the symbol ``$--$'' means that there is no adversarial network belonging to the \emph{weakest} category when the DALR strategy is applied.}
    \begin{tabular}{ccccccc}
    \toprule
    \textbf{network size } & \textbf{attack strategy} & \textbf{strong$\to$weak} & \textbf{strong$\to$weakest} & \textbf{strong$\to$super-weak} & \textbf{strong$\to$non-scale-free} & \textbf{overall} \\
    \midrule
    \multirow{3}[2]{*}{$n=500$} & \textbf{RLR} & 17.70\% & 15.60\% & 17.40\% & 5.40\% & 17.30\% \\
          & \textbf{DALR} & 6.40\% & \textbf{6.00\%} & 6.20\% & 5.40\% & 6.33\% \\
          & \textbf{DILR} & \textbf{3.40\%} & 7.80\% & \textbf{4.10\%} & \textbf{4.00\%} & \textbf{3.89\%} \\
    \midrule
    \multirow{3}[2]{*}{$n=1000$} & \textbf{RLR} & 17.00\% & 14.91\% & 16.72\% & 4.46\% & 16.19\% \\
          & \textbf{DALR} & 7.56\% & \textbf{9.60\%} & 7.16\% & 7.60\% & 7.68\% \\
          & \textbf{DILR} & \textbf{5.83\%} & 13.08\% & \textbf{4.47\%} & \textbf{3.95\%} & \textbf{5.51\%} \\
    \midrule
    \multirow{3}[2]{*}{$n=2000$} & \textbf{RLR} & 16.25\% & \textbf{16.38\%} & 16.78\% & \textbf{5.13\%} & 16.07\% \\
          & \textbf{DALR} & \textbf{8.28\%} & $--$     & 7.94\% & 7.60\% & \textbf{8.16\%} \\
          & \textbf{DILR} & 10.35\% & 20.20\% & \textbf{5.50\%} & 6.02\% & 8.71\% \\
    \bottomrule
    \end{tabular}%
  \label{effect}%
\end{table*}%

\begin{table*}[htbp]
  \centering
  \caption{The metrics of concealment ($\Delta{L}$, $\Delta{C}$ and $\Delta{D}$) when successfully attack on the networks of different sizes, for different attack strategies.}
    \begin{tabular}{cccccccc}
    \toprule
    \textbf{Metrics} & \textbf{network size } & \textbf{attack strategy} & \textbf{strong$\to$weak} & \textbf{strong$\to$weakest} & \textbf{strong$\to$super-weak} & \textbf{strong$\to$non-scale-free} & \textbf{overall} \\
    \midrule
    \multirow{9}[6]{*}{\textbf{$\Delta{L}$}} & \multirow{3}[2]{*}{$n=500$} & \textbf{RLR} & \textbf{2.28\%} & \textbf{2.37\%} & \textbf{2.32\%} & \textbf{1.33\%} & \textbf{2.25\%} \\
          &       & \textbf{DALR} & 10.03\% & 9.21\% & 9.86\% & 8.39\% & 9.96\% \\
          &       & \textbf{DILR} & 3.31\% & 6.58\% & 3.81\% & 4.10\% & 3.67\% \\
\cmidrule{2-8}          & \multirow{3}[2]{*}{$n=1000$} & \textbf{RLR} & \textbf{2.41\%} & \textbf{2.36\%} & \textbf{2.34\%} & \textbf{1.17\%} & \textbf{2.09\%} \\
          &       & \textbf{DALR} & 13.49\% & 17.40\% & 13.04\% & 12.92\% & 13.08\% \\
          &       & \textbf{DILR} & 5.72\% & 11.27\% & 4.64\% & 3.23\% & 5.07\% \\
\cmidrule{2-8}          & \multirow{3}[2]{*}{$n=2000$} & \textbf{RLR} & \textbf{2.10\%} & \textbf{2.02\%} & \textbf{2.06\%} & \textbf{0.81\%} & \textbf{2.07\%} \\
          &       & \textbf{DALR} & 16.23\% & $--$     & 15.86\% & 15.10\% & 16.08\% \\
          &       & \textbf{DILR} & 9.75\% & 16.29\% & 5.39\% & 3.50\% & 7.74\% \\
    \midrule
    \multirow{9}[6]{*}{\textbf{$\Delta{C}$}} & \multirow{3}[2]{*}{$n=500$} & \textbf{RLR} & 31.83\% & \textbf{29.30\%} & 31.89\% & \textbf{17.64\%} & 32.22\% \\
          &       & \textbf{DALR} & 60.28\% & 57.41\% & 61.47\% & 52.45\% & 60.90\% \\
          &       & \textbf{DILR} & \textbf{20.82\%} & 38.60\% & \textbf{23.50\%} & 26.80\% & \textbf{23.91\%} \\
\cmidrule{2-8}          & \multirow{3}[2]{*}{$n=1000$} & \textbf{RLR} & 31.74\% & \textbf{32.37\%} & 31.54\% & \textbf{13.35\%} & 30.11\% \\
          &       & \textbf{DALR} & 68.18\% & 87.09\% & 68.15\% & 64.53\% & 67.75\% \\
          &       & \textbf{DILR} & \textbf{28.96\%} & 53.53\% & \textbf{24.72\%} & 15.75\% & \textbf{26.62\%} \\
\cmidrule{2-8}          & \multirow{3}[2]{*}{$n=2000$} & \textbf{RLR} & \textbf{28.62\%} & \textbf{22.41\%} & 29.79\% & \textbf{9.61\%} & \textbf{31.56\%} \\
          &       & \textbf{DALR} & 71.69\% & $--$     & 71.76\% & 67.33\% & 72.85\% \\
          &       & \textbf{DILR} & 42.30\% & 64.47\% & \textbf{23.34\%} & 10.02\% & 35.24\% \\
    \midrule
    \multirow{9}[6]{*}{\textbf{$\Delta{D}$}} & \multirow{3}[2]{*}{$n=500$} & \textbf{RLR} & \textbf{0.00\%} & \textbf{0.08\%} & \textbf{0.00\%} & \textbf{0.11\%} & \textbf{0.05\%} \\
          &       & \textbf{DALR} & \textbf{0.00\%} & 0.95\% & 0.29\% & 0.59\% & 0.46\% \\
          &       & \textbf{DILR} & 0.02\% & 0.40\% & 0.06\% & 0.26\% & 0.06\% \\
\cmidrule{2-8}          & \multirow{3}[2]{*}{$n=1000$} & \textbf{RLR} & \textbf{0.00\%} & \textbf{0.16\%} & \textbf{0.01\%} & 0.31\% & 0.11\% \\
          &       & \textbf{DALR} & 0.01\% & 0.17\%\% & 0.03\% & 0.37\% & 0.13\% \\
          &       & \textbf{DILR} & \textbf{0.00\%} & 0.19\% & \textbf{0.01\%} & \textbf{0.04\%} & \textbf{0.04\%} \\
\cmidrule{2-8}          & \multirow{3}[2]{*}{$n=2000$} & \textbf{RLR} & \textbf{0.00\%} & 1.51\% & 0.04\% & \textbf{0.10\%} & 0.41\% \\
          &       & \textbf{DALR} & 0.03\% & $--$     & \textbf{0.02\%} & 0.13\% & \textbf{0.03\%} \\
          &       & \textbf{DILR} & 0.03\% & \textbf{0.25\%} & 0.05\% & 0.50\% & 0.17\% \\
    \bottomrule
    \end{tabular}%
  \label{tab:conceal}%
\end{table*}%

For the DILR attack strategy, there is a question how to divide the intervals according to the node degree. In general, the degree distribution of a scale-free network satisfies the Pareto principle, which has been widely used in sociology and business management, indicating that for a scale-free network the hub nodes are concentrated on the top 20\% of the degree ranking. For this reason, the $\gamma$ in the DILR strategy is set to be 20\%. However, because it is not clear how to determine the interval of intermediate node degrees, the indicator $\beta$ is tuned among five different values, 35\%, 40\%, 50\%, 60\%, and 70\%. The experimental results show that both the probability of a successful attack and the cost of the attack increase as $\beta$ increases. Considering the balance between the two, the indicator is set as $\beta=50\%$ in the simulation.

Now, the three attack strategies, RLR, DALR, and DILR, are applied onto the BA \emph{strong} scale-free networks of different sizes. An attack will be ended as soon as the adversarial network is not belong to the \emph{strong} category, no matter it is in \emph{weak}, \emph{weakest}, \emph{super-weak} or \emph{non-scale-free}. The following are three findings.

\begin{figure}[!t]
  \centering
  \includegraphics[scale=0.15]{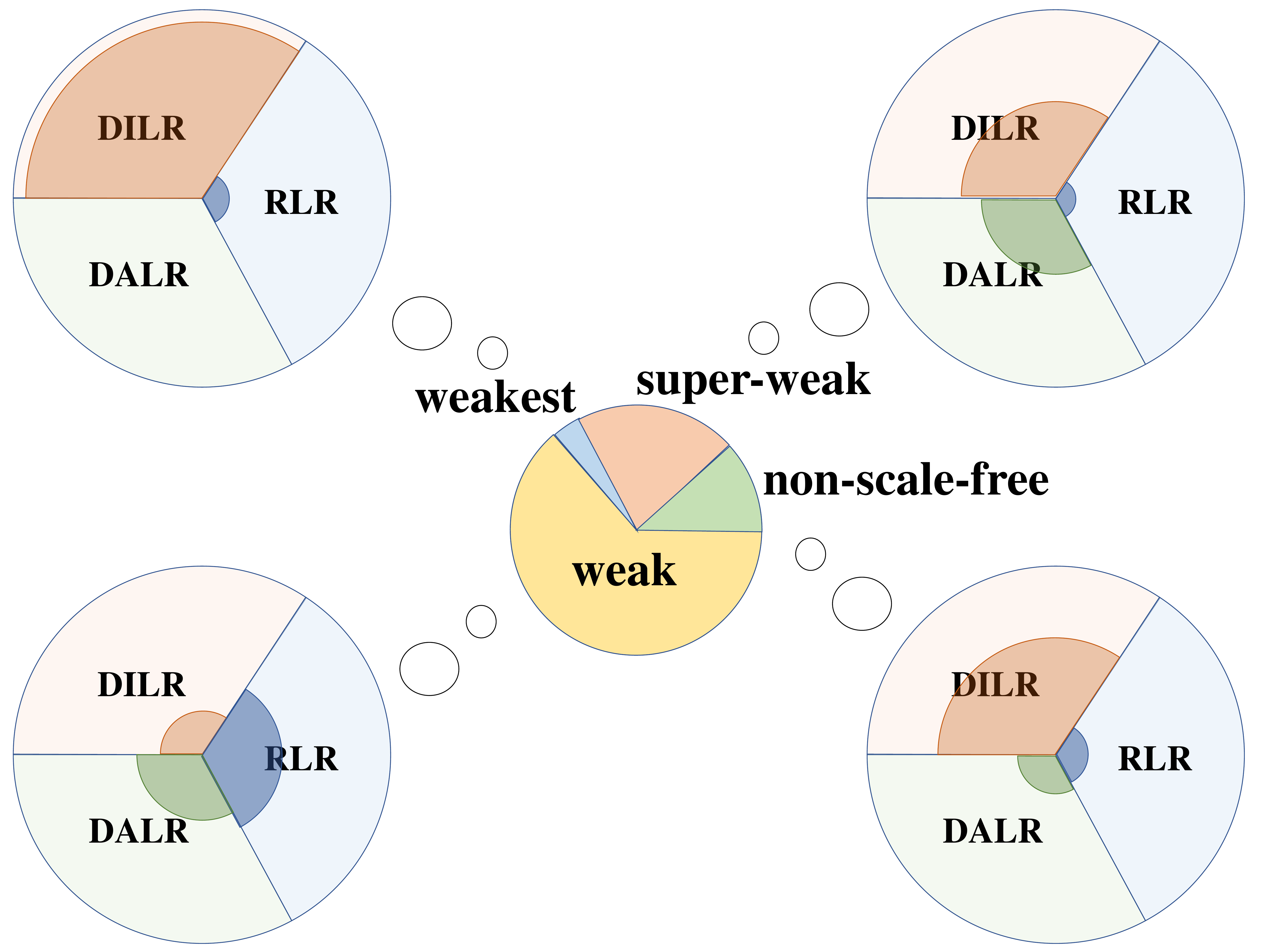}
  \caption{Probabilities of adversarial networks to belong to different categories. The middle pie chart represents the overall results by considering all the cases (three attack strategies on all the networks of different sizes). The surrounding pies represent the probabilities of the adversarial networks generated by different attack strategies for each category.}
  \label{fggailv}
\end{figure}

\begin{figure*}[!t]
  \centering
  \includegraphics[width=\linewidth]{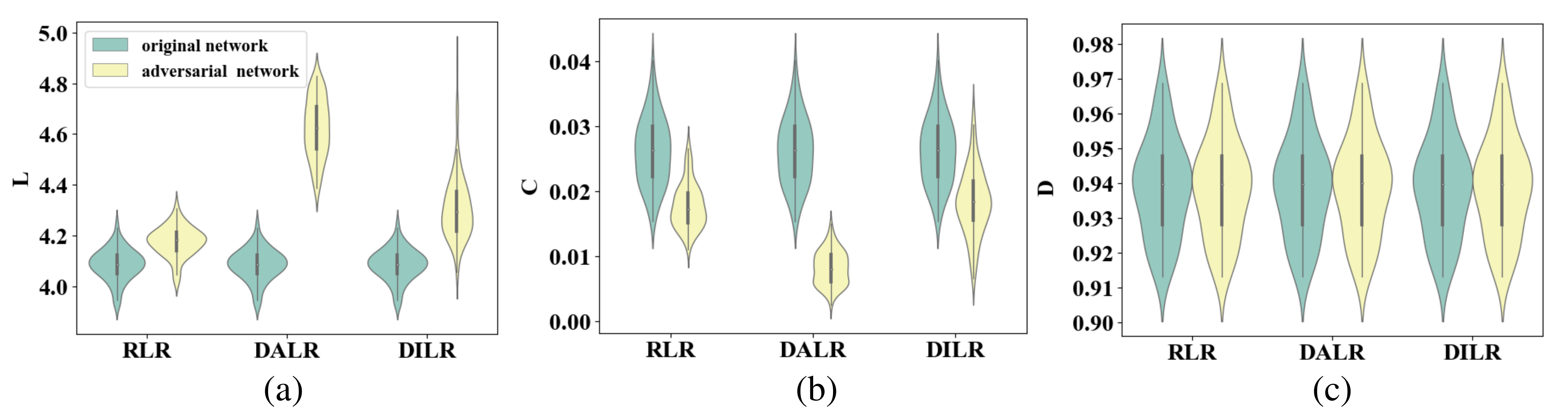}
  \caption{The violin plots of the three structural properties, $L$, $C$ and $D$, of the networks both before and after the attacks by different strategies.}
  \label{LCD}
\end{figure*}

\textit{First, it seems that the probabilities of the adversarial networks (obtained by attacking strong scale-free networks) belonging to different categories are quite different.} Overall, by considering all the cases (three attack strategies on the networks of different sizes), the adversarial networks are most likely to be in the weak category (with probability 62.92\%), but are most difficult to be in the weakest category (with probability 1.75\%), as shown in Fig.~\ref{fggailv}. By comparison, the adversarial networks are relatively easier to be in the super-weak category, rather than in the non-scale-free category, with properties 23.96\% and 11.37\%, respectively. This is because, generally, it is quite costly (a large number of links need to be rewired) to make $n_{tail}$ less than 50, especially for those networks of large sizes, where $n_{tail}\leq{50}$ is the threshold to define the weakest category. On the contrary, the other three parameters, $\alpha$, $R$ and $p$, are more sensitive to the attacks, since they are determined by the overall curve of the degree distribution, especially for the first two. Note that the fractions of adversarial networks in different categories are slightly different by attacks of different strategies, e.g., most of the adversarial networks in the weak category are generated by RLR, while most of those in the other three categories (weakest, super-weak, non-scale-free) are generated by DILR. But, overall, these strategies are consistent with each other.

\textit{Second, the two heuristic attack strategies are much more effective than the random rewiring strategy RLR, in the sense that a smaller $\Delta{M}$ is needed to succeed in the attack, while among them DILR is the most effective one.} As shown in TABLE~\ref{effect}, for RLR, which can be considered as random noise, typically more than 15\% links in the original networks need to be rewired to succeed in the attack (from strong to weak, weakest, or super-weak). However, surprisingly, only around 5\% links need to be rewired when the network is attacked to become non-scale-free, although such cases are rare. This result suggests that the BC classification could be robust against small random noise. More interestingly, when the perturbations are purposefully designed, the BC classification will be quite vulnerable, i.e., only 3.89\% links (38.9 in 1000 links on average) need to be rewired if DILR is applied to attack the networks of 500 nodes, much less than the value of 17.30\% by RLR. It seems that, as the network size increases, it is more difficult to succeed in an attack, e.g., the fractions of rewired links for both DALR and DILR steadily increases as the network size increases. Even so, the number of rewired links needed by DALR or DILR is only half of those needed by RLR, when the networks have 2000 nodes and 4000 links. By comparison, DILR is generally more effective than DALR, i.e., a smaller number of links are needed to succeed in attack when DILR is applied rather than DALR.

\textit{Third, the adversarial attack will not change the network structure by too much, i.e., the relative changes of $L$, $C$, and $D$ introduced by DILR are comparable with those introduced by RLR.} It is well known that the average shortest path-length $L$ and the average clustering coefficient $C$ are good for characterizing the global and local properties, respectively. On the other hand, the recently proposed diagonal distance $D$ is also a global property useful for detecting the dramatic changes in the topology of a network~\cite{tsiotas2019detecting}. Here, the relative changes of these three properties are used, i.e., $\Delta{L}$, $\Delta{C}$, and $\Delta{D}$, between the original networks and the adversarial networks, to measure if these properties will dramatically change under adversarial attacks. The results are presented in TABLE~\ref{tab:conceal}. It is found that, overall, $\Delta{L}$ introduced by DILR is larger than those introduced by RLR, but much smaller than that introduced by DALR, and that DILR and RLR have comparable $\Delta{C}$, around 30\%, which is only half of that introduced by DALR. All the three attack strategies introduce very small $\Delta{D}$, i.e., lower than 1\%. Note that, here $\Delta{C}$ is relatively large for all the three attack strategies, because BA scale-free networks do not have many triangular motifs, thus have very small average clustering coefficients. Consequently, a very small perturbation on a triangular structure will lead to a huge change of $\Delta{C}$ according to Eq.~(\ref{DeltaC}). To statistically compare the original networks and the adversarial networks generated by each attack strategy, the violin plots is shown for the three structural properties, i.e., $L$, $C$ and $D$, of the networks, both before and after an attack, as shown in Fig.~\ref{LCD}. It is found that the shortest path-length increases and the clustering coefficient decreases after the attack, while the diagonal distance remains almost the same. This is reasonable, since the attack focuses on destroying the \emph{scale-free} property, which will weaken the dominant role of hub nodes and further increase the shortest path-length. On the other hand, the global random rewiring operations in the three strategies may also destruct triangles in these networks, leading to smaller clustering coefficients. In general, DILR has comparable concealment as RLR, both of which introduce relatively small changes of $L$, $C$ and $D$; while DALR causes dramatic changes of the network structures, thus could be relatively easy to detect.

\section{Conclusion}\label{con}

This paper has discussed the robustness of the scale-free network classification method proposed by Broido and Clauset~\cite{broido2019scale}, referred to as BC classification herein. In so doing, three attack strategies, RLR, DALR and DILR, are proposed and tested. The attack experiments on strong scale-free networks generated by the BA model show that the BC classification method is vulnerable to adversarial attacks, e.g., only 6\% links need to be rewired to successfully attack a strong scale-free network such that it becomes a weak, weakest, super-weak, or even non-scale-free network. In addition, the BC classification result is not as good as described~\cite{broido2019scale}: The cost of transforming a strong network into the weakest category is even higher than the cost of transforming it into the non-scale-free category, which indicates that there is no layered relationship between these categories.

This paper has examined four indicators in the BC classification method: $p$, $n_{tail}$, $R$, and $\alpha$. Among these indicators, the values of $p$ and $\alpha$ can be easily changed by slightly disturbing the network structure, while $n_{tail}$ and $R$ show relatively strong robustness. This results in unbalanced distributions of the adversarial networks in different categories. For instance, since the requirement of the weakest category is $n_{tail}>50$, it is extremely difficult to attack a strong scale-free network to change it to be in the weakest category, especially when the network is relatively large. Therefore, one should consider the robustness for each indicator when trying to propose a new classification method.

This study may provide a different perspective to the arguable definition of scale-free networks. Moreover, beyond the scale-free feature, in complex networks there are also many other physical criteria derived from real-world data. Therefore, a suggestion is to adopt similar methods to test their robustness against various adversarial attacks.

\section*{Acknowledgments}

The authors would like to thank all the members in the IVSN Research Group, Zhejiang University of Technology, for their valuable discussions about the ideas and technical details presented in this paper.

\nocite{*}
\bibliography{reference}

\end{document}